\documentclass[sigconf]{acmart}

\AtBeginDocument{%
  }

\setcopyright{acmlicensed}
\copyrightyear{2025}
\acmYear{2025}
\setcopyright{cc}
\acmConference[UMAP Adjunct '25]{Adjunct Proceedings of the 33rd ACM
Conference on User Modeling, Adaptation and Personalization}{June 16--19,
2025}{New York City, NY, USA}
\acmBooktitle{Adjunct Proceedings of the 33rd ACM Conference on User
Modeling, Adaptation and Personalization (UMAP Adjunct '25), June 16--19,
2025, New York City, NY, USA}
\acmDOI{10.1145/3708319.3734176}
\acmISBN{979-8-4007-1399-6/2025/06}


\graphicspath{ {images} }

\begin{document}

\title{Hybrid Personalization Using Declarative and Procedural Memory Modules of the Cognitive Architecture ACT-R}

\iffalse
\author{Kevin Innerebner}
%\authornote{Both authors contributed equally to this research.}
\orcid{0009-0001-8556-5371}
\affiliation{%
  \institution{Graz University of Technology}
  \city{Graz}
  %\state{}
  \country{Austria}
}
\email{innerebner@tugraz.at}

\author{Dominik Kowald}
\orcid{0000-0003-3230-6234}
\affiliation{%
  \institution{Know Center Research GmbH \& Graz University of Technology}
  \city{Graz}
  \country{Austria}
}
\email{dkowald@know-center.at}

\author{Markus Schedl}
\orcid{0000-0003-1706-3406}
\affiliation{%
  \institution{Johannes Kepler University Linz \& LIT AI Lab}
  %\institution{Linz Institute of Technology AI Lab}
  \city{Linz}
  \country{Austria}
}
\email{markus.schedl@jku.at}

\author{Elisabeth Lex}
\orcid{0000-0001-5293-2967}
\affiliation{%
  \institution{Graz University of Technology}
  \city{Graz}
  \country{Austria}
}
\email{elisabeth.lex@tugraz.at}
\fi

\author{Kevin Innerebner}
%\authornote{Both authors contributed equally to this research.}
\orcid{0009-0001-8556-5371}
\affiliation{%
  \institution{Graz University of Technology}
  \department{Institute of Human-Centred Computing}
  \city{Graz}
  %\state{}
  \country{Austria}
}
\email{innerebner@tugraz.at}

\author{Dominik Kowald}
\orcid{0000-0003-3230-6234}
\affiliation{%
  \institution{Know Center Research GmbH}% \& Graz University of Technology}
  \department{FAIR-AI}
  \city{Graz}
  \country{Austria}
}
\affiliation{%
  \institution{Graz University of Technology}
  \department{Institute of Human-Centred Computing}
  \city{Graz}
  \country{Austria}
}
\email{dkowald@know-center.at}

\author{Markus Schedl}
\orcid{0000-0003-1706-3406}
\affiliation{%
  \institution{Johannes Kepler University Linz}% \& LIT AI Lab}
  %\institution{Linz Institute of Technology AI Lab}
  \department{Institute of Computational Perception}
  \city{Linz}
  \country{Austria}
}
\affiliation{%
  \institution{Linz Institute of Technology AI Lab}
  \department{Human-centered Artifical Intelligence}
  \city{Linz}
  \country{Austria}
}
\email{markus.schedl@jku.at}

\author{Elisabeth Lex}
\orcid{0000-0001-5293-2967}
\affiliation{%
  \institution{Graz University of Technology}
  \department{Institute of Human-Centred Computing}
  \city{Graz}
  %\state{}
  \country{Austria}
}
\email{elisabeth.lex@tugraz.at}

\renewcommand{\shortauthors}{Innerebner, Kowald, Schedl, Lex}

\begin{abstract}
Recommender systems often rely on sub-symbolic machine learning approaches that operate as opaque black boxes. These approaches typically fail to account for the cognitive processes that shape user preferences and decision-making. In this vision paper, we propose a hybrid user modeling framework based on the cognitive architecture ACT-R that integrates symbolic and sub-symbolic representations of human memory. Our goal is to combine ACT-R's declarative memory, which is responsible for storing symbolic chunks along sub-symbolic activations, with its procedural memory, which contains symbolic production rules. This integration will help simulate how users retrieve past experiences and apply decision-making strategies. With this approach, we aim to provide more transparent recommendations, enable rule-based explanations, and facilitate the modeling of cognitive biases. We argue that our approach has the potential to inform the design of a new generation of human-centered, psychology-informed recommender systems.
\end{abstract}


\maketitle

\section{Introduction}
Recommender systems are essential in today's digital ecosystems, yet most rely on opaque, sub-symbolic models such as deep neural networks. These approaches lack transparency and often do not account for cognitive processes underlying human decision-making behavior~\cite{lex2021psychology,schedl2024trustworthy}. 
Hybrid artificial intelligence (AI) systems that integrate symbolic reasoning with sub-symbolic learning offer a promising alternative~\cite{colelough2025neuro} to advance human-centric personalization~\cite{spillo2024recommender}. This mirrors dual-process theories in cognitive psychology, which distinguish fast, intuitive processes (i.e., System 1) from slower, deliberative reasoning (i.e., System 2)~\cite{kahneman2011thinking}. However, many recommender systems do not incorporate these cognitive theories. 
%While sub-symbolic models capture the former very well (e.g., pattern recognition), symbolic methods can also model structured reasoning processes. Bringing these two approaches together holds the promise of developing personalized recommender systems that not only learn from user data but also reason about user preferences and goals.   
%A natural extension of this paradigm can be found in the field of psychology-informed recommender systems, which draws on insights from cognitive science and psychology to model and predict user behavior~\cite{tkalvcivc2024inferring,curmei2022towards,www_sustain,winecoff2019users,lex2021psychology,tkalvcivc2020complementing}. In particular, 

Cognitive architectures such as ACT-R (Adaptive Control of Thought Rationale)~\cite{anderson2004integrated} offer a psychologically grounded hybrid framework that aligns with these theories. ACT-R is a computational framework that simulates human cognition, such as perception, memory, attention, learning, and decision-making~\cite{kotseruba202040}, via interacting memory modules. Its declarative memory models memory retrieval using symbolic chunks and sub-symbolic activation, while its procedural memory governs symbolic decision making via production rules. Production rules in ACT-R are \texttt{IF-THEN} condition-action pairs that determine user behavior.

In prior work in recommender systems, ACT-R's declarative memory has been used to model recency and frequency effects in user behavior~\cite{tran2024transformers,moscati2023integrating,reiter2021predicting,lex2020modeling,ismir_lfm_2019,kowald2015evaluating,kowald2017temporal,kowald2024transparent}. However, the symbolic reasoning capabilities of ACT-R's procedural memory remain largely untapped, except for attempts in somewhat related fields like Web navigation~\cite{pirolli_snif-act_2003,fu2007snif}. 

In this paper, we propose a hybrid framework integrating ACT-R's declarative and procedural memory modules to better capture human decision-making and enable transparent personalization. 

% We argue that this integration allows for improved personalization and explainability, and user-specific modeling of cognitive biases. We outline this vision by (i) reviewing related work, (ii) introducing our modeling framework, and (iii) proposing future research directions. 

%rather than focusing solely on memory-based retrieval of prior interactions, our approach also emphasizes the symbolic reasoning processes that shape user behavior. We outline our envisioned approach in the next section.

%%%%%%%%%%%%%%%%%%%%%%%%%%%%%%%%%%%%%%%%%%%
\section{Background: ACT-R Architecture}
ACT-R (i.e., Adaptive Control Thought Rationale) is a modular cognitive architecture designed to simulate human cognition via interacting memory modules~\cite{anderson2004integrated}. Its hybrid architecture supports sub-symbolic mechanisms, such as knowledge retrieval from declarative memory, and symbolic processes encoded as procedural rules in procedural memory~\cite{anderson2004integrated}. The declarative memory stores knowledge in the form of chunks, and simulates human-like knowledge retrieval through ACT-R's activation equation, while the activation of a chunk is influenced by recency, frequency, and the current context in which the knowledge is used. The procedural memory governs behavior through symbolic production rules, i.e., symbolic \texttt{IF-THEN} statements, representing decision-making strategies.

% TODO: shorten and rewrite the below paragraph
%The declarative memory activation is computed from multiple modules. The modules are equations, which are inspired by the human cognitive retrieval process, and together compute the activations for retrieval. For example, the Base-Level Learning equation~\cite{anderson2004integrated}---one of the most commonly used modules---models temporal dynamics of chunk activations as follows:
% TODO: interpretation of BLL component
% TODO: also add softmax version (Markus paper) to get "probabilities"

%\subsection{Activation Equation}
One main component of ACT-R's declarative memory module is the activation equation.
Here, each user interaction can be represented as a chunk. The activation \(A_i\) of a chunk \(i\) for a user \(u\) is computed using the activation equation~\cite{anderson2004integrated}:
\begin{equation} \label{eq:A}
    A_i~=~B_i + \sum_j{W_j \cdot S_{j,i}}
\end{equation}
where \(B_i\) is the \textit{base-level activation}, which denotes the relevance of chunk \(i\) over time, based on the frequency and recency of its prior use, computed according to the \textit{power law of forgetting}, which models how the activation of memory traces decay over time~\cite{anderson1991reflections}:
\begin{equation} \label{eq:bll}
    B_i~=~ln\left(\sum\limits_{j~=~1}\limits^{n}{t_{j}^{-d}}\right)
  \end{equation}
In this equation, \(n\) denotes the number of prior occurrence (i.e., frequency) of chunk \(i\), \(t_j\) indicates the time since the $j$\textsuperscript{th} occurrence, and \(d\) is a decay parameter that models the rate at which information becomes less accessible in memory. The exponent \(t_{j}^{-d}\) formalizes the power law of forgetting. 

The second part of Equation~\ref{eq:A}, i.e., \(\sum_j{W_j \cdot S_{j,i}}\), accounts for the associative activation of chunk \(i\), which reflects its contextual relevance given the current state or focus of the user. The underlying intuition is that a chunk is more likely to be retrieved if it has strong semantic or episodic connections to what the user is currently doing or has recently interacted with~\cite{anderson1983spreading}. This component models how strongly chunk \(i\) is associated with other chunks \(j\) that are currently active in working memory or the knowledge retrieval context; \(W_j\) denotes the attentional weight of context element \(j\) and \(S_{j,i}\) the strength of association between chunk \(j\) and chunk \(i\), typically computed based on co-occurrence frequencies. 

\section{Our Vision: Hybrid Modeling Approach}

\begin{figure*}
    \centering
    \includegraphics[width=0.8\linewidth]{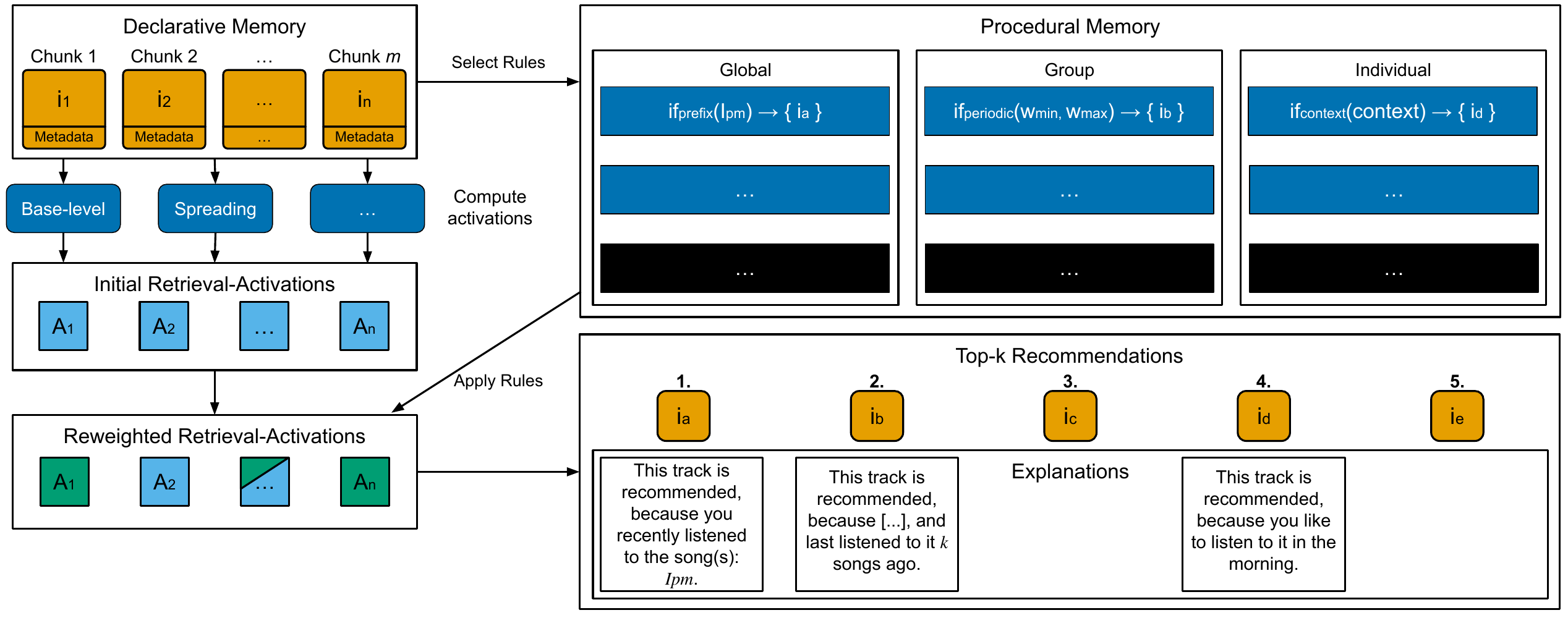}
    \caption{Visualization of our envisioned hybrid modeling approach based on ACT-R. The declarative memory stores item interactions as chunks and computes initial activations. The \texttt{IF} conditions of the production rules in the procedural memory are checked against these activations and could lead to reweighted activations. The items associated with the highest reweighted activations form our top-$k$ recommendations, for which explanations are provided based on the applied production rules.} 
    %\Description{A visualization of the declarative memory and procedural memory modules of the ACT-R architecture. We illustrate how these modules compute activations for chunks.}
    \label{fig:approach}
\end{figure*}

We propose a hybrid framework, depicted in Figure~\ref{fig:approach}, that integrates memory-based retrieval from ACT-R's declarative memory with rule-based decision strategies from its procedural memory. In our framework, chunks in declarative memory represent previously interacted items, such as songs or articles. While these chunks primarily encode item-level information, they may also include contextual metadata (e.g., time of day) to support context-aware rule reasoning. The system computes each chunk's activation based on its recency, frequency, and relevance to the current context. The system uses base-level activation (see Equation~\ref{eq:bll}) and optional extensions (e.g., spreading, partial matching) to compute relevance. The most highly activated chunks form the basis for recommendations. For instance, in a music recommendation scenario, a chunk may represent a previously played track, and its activation reflects how likely the user is to want to listen to it or a similar track next.

Additionally, production rules are applied that modify chunk activations based on symbolic reasoning. These rules model behavioral patterns such as novelty seeking, daily routines (e.g., morning playlists), or preference shifts, and are triggered by factors such as item sequences, recurring behaviors, or contextual cues.

\subsection{Production Rules}
\label{ssec:production-rules}
Production rules are symbolic \texttt{IF-THEN} structures that simulate how users evaluate recommendations, apply heuristics, and pursue goals. Such rules can be informed by psychological theory, derived from interaction data, or learned through user studies. In the context of recommender systems, we use them to simulate user strategies or heuristics, such as preferences for novelty, recurring listening patterns, or contextual behavior (e.g., time-of-the-day behavior). For example, \texttt{IF} the user listened to a track from an album \texttt{THEN} boost the next track of the album. Or \texttt{IF} the user frequently listens to a song every morning, \texttt{THEN} increase its activation in the morning context.
When the condition of a rule is met, it modifies the activation of one or more chunks, thereby influencing which recommendations are generated for the user. A straightforward implementation could be additive:
\begin{equation} \label{eq:add}
    A_k \leftarrow A_k + v_r, \quad \text{for each } i_k \in C_r
\end{equation}
where $v_r$ is the influence weight of the rule and $C_r$ is the list of chunks to boost.

\subsubsection{Classes of Rule Conditions}
We define three broad classes of rule conditions, all of which can be extracted from user interactions:

\begin{itemize}
%\textit{Sequential (Prefix) Rules:}
    \item \textbf{Sequential (Prefix) Rules:} These rules are triggered when a sequence of items occurs in the user's interaction history. For instance, if a user listens to tracks \texttt{A → B → C}, the system may learn a rule to prioritize track \texttt{D} next. Formally, we can represent a user's interaction history as an ordered list of item chunks %:
    %$I_{pm} \subseteq C_{n-w,n}$. 
    %\begin{equation}
    %    C = \{i_1,i_2,....,i_n\}, i_k \in I
    %\end{equation}
    \begin{equation}
        C = (i_1,i_2,....,i_n), i_k \in I,
    \end{equation}
    where $I$ is the set of all possible items, and items may be duplicated (i.e., $i_k$ may equal $i_{l}$, for $k\neq l$).
    A prefix rule fires when a known subsequence \( I_{pm} \subseteq C_{n-w,n} \) is detected within a lookback window \( w \). When triggered, it boosts the activation of predicted items. To mine sequential rules, we aim to experiment with sequential rule mining algorithms, such as RuleGrowth~\cite{DBLP:conf/sac/Fournier-VigerNT11} to identify rules of the form \(X \rightarrow Y\), where \(X\) and \(Y\) are unordered sets of items. If \(X\) appears in a user's interaction history, we can boost the activation of item chunks matching \(Y\), formally defined as $C_r=(i_k \in C \mid i_k \in Y)$ (e.g., using Equation~\ref{eq:add})
    %\(Y\) (e.g., using Equation~\ref{eq:add}).
    
%\textit{Periodic Rules:}
    \item \textbf{Periodic Rules:} These rules capture periodic occurrences of items, such as a user listening to their favorite song every few songs. We can represent this by identifying items that are repeatedly consumed within an item-based window:
    %$i_p\in C_{n - w_{max},n-w_{min}}$.
    \begin{equation}
        i_p\in C_{n - w_{max},n-w_{min}}
    \end{equation}
    Here, \( w_{\text{min}} \) and \( w_{\text{max}} \) denote the interval in which items reoccur. To extract periodic rules, we plan to explore periodic pattern mining algorithms, such as PFPM~\cite{Fournier-Viger17}. 
    
%\textit{Context-aware Rules:} 
    \item \textbf{Context-aware Rules:} These rules incorporate external signals such as time, day, or device. To identify context-aware rules, we can extract co-occurrences between items and metadata. Furthermore, we plan to leverage time series analysis to identify seasonality effects, such as listening to festive music during the winter holidays. To extract time-aware rules, we aim to investigate \emph{partial} periodic pattern~\cite{DBLP:conf/edbt/KiranSTK15} mining algorithms, such as RP-growth~\cite{DBLP:conf/edbt/KiranSTK15} or LPP-growth~\cite{DBLP:journals/isci/Fournier-VigerY21}.
\end{itemize}

\subsubsection{Rule Adaptation}
In our proposed framework, production rules can be adapted to different levels of granularity, enabling a tradeoff between personalization and generalization. At the individual level, rules are derived from a user's behavior, in line with ACT-R as a cognitive model of individual behavior. Such rules correspond to personal preferences, goals, and individual strategies. At the same time, rules can be aggregated across groups of users with similar behavior and characteristics, such as user demographics, cultural background, or shared preferences (e.g., a group's preference for music from their home country). Corresponding group-level rules allow for capturing patterns that are more specific than global trends. Finally, we could also learn \textit{global} production rules from all users' data. Such rules could capture general trends or widely shared preferences. Thus, these rules can serve as a fallback when user- or group-specific data is missing. Critically, we believe that rules at the individual and group levels enable us to address important challenges in personalization, such as accounting for users with non-mainstream behavior~\cite{kowald2023study,kowald2020unfairness,bauer2019global}.

\subsection{Explainability}
A key advantage of our envisioned hybrid framework is its inherent transparency. Since production rules are symbolic and influence item activations, they enable the generation of explanations for recommendations in terms of interpretable \texttt{IF-THEN} rules. When a production rule contributes to the activation of a chunk, we can generate a corresponding explanation reflecting the specific condition that triggered the production rule. In this way, we can offer users deeper insights into the \textit{``Why''} behind a recommendation~\cite{tintarev2007explanations}. For example, in a music recommender, we can generate explanations depending on the \texttt{IF} condition of the applied rule (see Section~\ref{ssec:production-rules}):
\begin{itemize}
    \item \textbf{Prefix Matching:} \textit{``This track is recommended, because you recently listened to the song(s): $I_{pm}$.''}
    \item \textbf{Periodic Matching:} \textit{``This track is recommended, because you regularly listen to this song, and listened to it $k$ songs ago.''}
    \item \textbf{Context-aware Matching:} \textit{``This track is recommended, because you like to listen to it in the morning.''} 
\end{itemize}
Furthermore, the granularity of the rule, i.e., whether it is \emph{global}, \emph{user-group}, or \emph{individual}, could also be presented to the user. This further provides users with information on how the system categorizes them, at the cost of reduced privacy.
Additionally, the system could be extended to be interactive and allow users to disable rules that they find unfitting, thereby tuning the system to their needs, without requiring deep technical knowledge. 
%Options for the user to and provides the opportunity for the user to provide feedback and tune the recommendations 
Finally, the symbolic nature of our rules facilitates counterfactual analysis. For example, by disabling particular production rules, one can evaluate the impact of specific production rules on recommendation outcomes. Such functionality can support users, system designers, and system providers in identifying which rules enhance user satisfaction and which may introduce unintended biases or effects.
% TODO: could be mentioned, but as LLM outputs aren't explainable themselves, it's hard to argue.
%Another opportunity to improve the explainability is to use LLMs to create individual explanations for rules. For example, if a \emph{prefix matching} rule was mined that suggests song A is often followed by song B, providing metadata about both songs, such as artist, title, lyrics, tempo etc., might allow an LLM to provide a more in-depth explanation.

%Because the production rules are symbolic, the applied rules can be provided as explanations along with the recommendations, and enhance transparency.

%While the declarative memory module uses a subsymbolic, activation-based retrieval, which is not 

%While the declarative memory module computes subsymbolic activations and is not directly explainable, they are a simple linear combination of modules, for which the intuition can be explained. For example, the Base-Level Learning equation balances recency and frequency, while the spreading module activates for items that co-occur with the previous item the user interacted with. Nevertheless, this 
%This would allow for explainable user personalization and provides the option to group users based on user-specific production rules.

\subsection{Cognitive Biases}
While ACT-R is a model of rational human cognition, it can also provide a foundation for studying deviations from rationality, i.e.,~irrational behavior.
Such deviation of the individual from rationality and objectivity in terms of judgment or decision making is commonly referred to as \textit{cognitive bias}.
Cognitive biases often happen unconsciously in humans and have been studied in psychology, sociology, and behavioral economics for decades~\cite{dobelli2013art,kahneman2011thinking}.
While some of the plethora of cognitive biases from psychological and sociological literature have also been observed in recommender system environments, e.g.,~position bias, anchoring, bandwagon effect or popularity bias, confirmation bias, and conformity bias~\cite{lex2021psychology}, research to make them transparent and mitigate their negative effects (e.g., unfairness of popularity bias) is still in its early stages.

Our envisioned hybrid framework offers a promising avenue for operationalizing cognitive biases. Since production rules in ACT-R explicitly model decision strategies, they can also be adapted to counteract undesirable behavioral patterns or algorithmic behavior. 

We propose to formulate rules manually or semi-automatically that counteract certain cognitive biases. Examples may include rules to mitigate conformity bias, e.g., by giving higher weights to more diverse content or content that is not in line with the user's common preferences based on item-side information. 
Another example could be rules to consider the popularity of items in a user's interactions and adjust the popularity level of recommendations to counteract popularity bias. 
Yet another approach could involve defining rules that reduce exaggerated cultural homophily between content creators and consumers~\cite{DBLP:conf/intrs/SchedlLM24}.

%In sociology, they typically refer to collective prejudices of a society that favor one group's values, norms, and traditions over others~\cite{eberhardt2020biased,bonilla2006racism}.

%%%%%%%%%%%%%%%%%%%%%%%%%%%%%%%%%%%%%%%%%%
\section{Conclusions and Future Work}
%We presented a vision for a hybrid user modeling framework based on the cognitive architecture ACT-R that integrates declarative and procedural memory components. We outlined the potential of our framework to generate transparent personalized recommendations because of the explicit structure of rule-based reasoning. Furthermore, the framework will allow for the adaptation of rules at different levels---i.e., individual, group, or global---and provides a basis for incorporating interventions designed to study cognitive biases. 
 We introduced a vision for a hybrid user modeling framework that integrates ACT-R’s declarative and procedural memory modules to generate transparent, cognitively grounded recommendations. By combining activation-based retrieval with symbolic rule reasoning, our approach enables transparent recommendations and supports the study of cognitive biases.

In future work, we plan to systematically investigate methods for learning such production rules. We also aim to explore the use of large language models (LLMs) to support rule generation and refinement.
For evaluation,  we will consider dimensions such as cognitive plausibility, transparency, fairness, and user satisfaction, by using, e.g., user studies and online evaluation methodologies. Additionally, we plan to use counterfactual analysis to assess the effects of individual rules on recommendation outcomes.

\begin{acks}
Funding sources: Austrian Science Fund (FWF): 10.55776/COE12, Cluster of Excellence {\textcolor{blue}{\href{https://www.bilateral-ai.net/home}{Bilateral Artificial Intelligence}}}, Austrian FFG COMET program, FFG HybridAir project: \#FO999902654.
\end{acks}

\bibliographystyle{ACM-Reference-Format}
\bibliography{main}

%%
%% If your work has an appendix, this is the place to put it.
%\appendix

%\section{Research Methods}

\end{document}